\documentclass[aps,prd,twocolumn,superscriptaddress,nofootinbib,showpacs,10pt]{revtex4-1}
\usepackage[utf8]{inputenc}
\usepackage[english]{babel}
\usepackage{amsmath}
\usepackage{amsfonts}
\usepackage{amssymb}
\usepackage{graphics}
\usepackage{graphicx}
\usepackage[left=2cm,right=2cm,top=2cm,bottom=2cm]{geometry}
\usepackage[usenames,dvipsnames,svgnames]{xcolor}  
\usepackage{subfigure}
\usepackage{hyperref}   
\hypersetup{colorlinks=true,linkcolor=beamer@PRD, citecolor=beamer@PRD}
\definecolor{beamer@PRD}{RGB}{46,48,146}
\newcommand\myref[1]{\textcolor{beamer@PRD}{(}\ref{#1}\textcolor{beamer@PRD}{)}}

\usepackage{todonotes}
\usepackage{braket}
\begin{document}
\title{Controlling decoherence via $\mathcal{PT}$-symmetric non-Hermitian open quantum systems}
\author{Sanjib Dey}\email{dey@iisermohali.ac.in}
\affiliation{Department of Physical Sciences, Indian Institute of Science Education and  Research Mohali, \protect\\ Sector 81, SAS Nagar, Manauli 140306, India}
\author{Aswathy Raj}
\affiliation{Department of Physical Sciences, Indian Institute of Science Education and  Research Mohali, \protect\\ Sector 81, SAS Nagar, Manauli 140306, India}
\author{Sandeep K.~Goyal}
\affiliation{Department of Physical Sciences, Indian Institute of Science Education and  Research Mohali, \protect\\ Sector 81, SAS Nagar, Manauli 140306, India}
\begin{abstract}
We have studied the effect of a non-Hermitian Bosonic bath on the dynamics of a two-level spin system. The non-Hermitian Hamiltonian of the bath is chosen such that it converges to the harmonic oscillator Hamiltonian when the non-Hermiticity is switched off. We calculate the dynamics of the spin system and found that the non-Hermiticity can have positive as well as negative effects on the coherence of the system. However, the decoherence can be completely eliminated by choosing the non-Hermiticity parameter and the phase of the system bath interaction appropriately. We have also studied the effect of this bath on the entanglement of a two-spin system when the bath is acting only on one spin.    
\end{abstract}


\maketitle
\section{Introduction} \label{sec1}
Loss of information and coherence due to the interaction of a quantum system with its surrounding is familiar as decoherence~\cite{Leggett_etal,Breuer_Book}. The same phenomenon is responsible for emergence of classicality in all the quantum systems around us~\cite{Schlosshauer_Book,Plenio_Knight}. In order to make the quantum computing platforms feasible and commercial we need to be able to control the decoherence. There have been a lot of investigations to prevent the loss of coherence of quantum systems arising from different scenarios; see, for instance \cite{Vitali_Tombesi_Milburn, Horoshko_Kilin, Viola_Lloyd, Thorwart_etal, Agarwal_Scully_Walther, Branderhorst_etal, Poletti_etal}. Besides the standard quantum mechanical framework, $q$-deformed models \cite{Dehdashti_etal} have also been utilized for the purpose. Here we present a model in which a two-level quantum system (spin) interacts with a large bath of harmonic oscillators where the dynamics of the bath is governed by non-Hermitian Hamiltonians. We show that the non-Hermitian character of the bath Hamiltonian gives rise to suppression of the decoherence effects; hence long coherence time. Furthermore, the non-Hermitian bath provides us additional parameters which can be exploited to control the decoherence.

Traditional quantum mechanics considers only Hermitian operators as bona fide physical observables because they exhibit real spectrum. The operators which are invariant under simultaneous parity and time inversion are also known to possess real eigenvalues~\cite{Bender_Boettcher,PT_Book}. This has given rise to a new field called non-Hermitian quantum systems. The phenomena of pseudo and quasi-Hermitian Hamiltonian systems also play important roles in the formulation of such systems \cite{Scholtz_Geyer_Hahne,Mostafazadeh}. Most credit for the fame of this field can be attributed to the measurement problem in traditional quantum mechanics which demands for a modification in the very foundations of the laws of quantum mechanics. Apart from various theoretical studies \cite{Znojil,Klaiman,Longhi,Graefe_Jones,Cartarius,Dey_Fring_Mathanaranjan,Dey_Fring_Gouba}, researchers have performed plenty of experiments and found the existence of non-Hermiticity in several physical systems, most notably, in optics \cite{Guo,Ruter}, photonics \cite{Regensburger,Feng}, acoustics \cite{Fleury,Shi}, quantum walk \cite{Xiao}, material sceience \cite{Weimann}, microresonator \cite{Chang,Peng}, etc.

In this article, we exploit the non-Hermitian formalism to devise methods to control the decoherence in a single two-level quantum system that are interacting with a large Bosonic bath of harmonic oscillators. The interaction between the bath and the system is chosen to be of pure dephasing type. Therefore, the effect of the interaction with the bath will be felt by only the off-diagonal terms of the system density matrix while the populations of the system energy levels will be unaffected. We choose the bath Hamiltonian to be non-Hermitian in nature  which depends on a dimensionless parameter $\tau$. In the limit $\tau \to 0$ the bath Hamiltonian acquires the form of harmonic oscillator Hamiltonian.

We show that by choosing the non-Hermiticity parameter $\tau$ appropriately we can suppress the decoherence to almost zero.  Moreover, we indicate how it may be possible to obtain maximum robustness in decoherence from our model. Another interesting finding of our work is that in the presence of non-Hermitian Hamiltonians the decoherence factor not only depends on the strength of the interaction between the system and the bath but also on the phase of the coupling strength. We show the decoherence characteristics of the spin-system by studying the entanglement of a two-spin system where the initial state of the system is a maximally entangled state and one of the spin is interacting with the bath. Interestingly, we found that the entanglement in our system persists indefinitely in the entire $\mathcal{PT}$-symmetric regime, unlike the results shown in~\cite{Fring_Frith,Chakraborty}, albeit in different settings, which only hold within the spontaneous $\mathcal{PT}$-broken regime and on the exceptional points, respectively.
 
The article is organized as follows: In Sec.\,\ref{sec2}, we introduce some prerequisite in order to set up the framework for analyzing the decoherence properties and entanglement in non-Hermitian systems. We, then, apply these concepts to a non-Hermitian open quantum system model in Sec.\,\ref{sec3}, where we show the system dynamics in the presence of non-Hermitian bath. We conclude in Sec.\,\ref{sec4}.
\section{Background}\label{sec2}
In this section, we present the relevant background of the Open quantum system dynamics and the non-Hermitian Hamiltonian. We also introduce the entanglement measure which is used to calculate the entanglement in the system interacting with non-Hermitian bath.  
\subsection{Dynamics of open quantum systems} \label{sec2a}
An open quantum system usually refers to a combined model consisting of two quantum bodies, the system $\mathcal{S}$ (living in a Hilbert space $\mathcal{H}_{\text{S}}$) and the bath $\mathcal{B}$ (living in another Hilbert space $\mathcal{H}_{\text{B}}$), such that the entire system $\mathcal{S+B}$ lives in the composite space $\mathcal{H}=\mathcal{H}_{\text{S}}\otimes \mathcal{H}_{\text{B}}$. The total system $\mathcal{S+B}$ is considered isolated, which results in the dynamics of the total system to be unitary. However, the individual subsystems $\mathcal{S}$ and $\mathcal{B}$ are not isolated to each other, but their mutual interaction causes a non-unitary dynamics for both of them. This non-unitary dynamics, the reduced dynamics, of the system due to its interaction with the bath results in decoherence and eventually classicality~\cite{Breuer_Book}. 

The reduced dynamics of the system $\mathcal{S}$ can be obtained by first evolving the total system plus bath using the unitary dynamics and then eliminating the bath degrees of freedom. For simplicity, let us suppose that at time $t=0$, the system and bath are uncorrelated, i.~e., $\rho_{\text{SB}}(0)=\rho_{\text{S}}(0)\otimes \rho_{\text{B}}$, where $\rho_{\text{S}}(0)$ is the density matrix of the system at time $t=0$ and $\rho_{\text{B}}$ is the steady state of the bath.  The state of the total system at time $t$  reads $\rho_{\text{SB}}(t)=U_{\text{SB}}(t)\rho_{\text{SB}}(0)U_{\text{SB}}^\dagger(t)$, where $U_{\text{SB}}(t)$ is the time propagator for the total system. Since, we are interested only in the dynamics of the system $\mathcal{S}$ we can trace out the bath which results in
\begin{equation}
\rho_{\text{S}}(t)=\text{Tr}_{\text{B}}\left\{\rho_{\text{SB}}(t)\right\}.
\end{equation}
We consider a simple open quantum system in which a two-level quantum system (spin) is interacting with a bath of infinitely many harmonic oscillators~\cite{Leggett_etal}. The interaction of the spin with the bath is such that the free spin Hamiltonian commutes with the total Hamiltonian of the spin plus bath. Therefore, the interaction of the spin with the bath does not affect the populations of the different spin energy levels. The Hamiltonian for such a system reads
\begin{equation}\label{HamSB}
H= \frac{\omega_0}{2}\sigma_z + \sum_{k}^{}\omega_ka_k^\dagger a_k +\sigma_z\otimes \sum_{k}^{}(g_{k}a_k^\dagger+g_k^\ast a_k).
\end{equation}
Here, $\sigma_z$ is the Pauli matrix along the $z$-axis, and $a_k$ and $a_k^\dagger$ are the creation and annihilation operators for the $k$-th mode of the Harmonic oscillator. The complex number $g_k$ characterizes the interaction strength between the spin and the $k$-th mode of the harmonic oscillator. Here, $\omega_0$ and $\omega_k$ are the natural frequencies of the spin and the harmonic oscillators. For simplicity, we have considered $\hbar=1$.

The calculations of the system dynamics can be simplified by going to the interaction picture. The interaction Hamiltonian in the interaction picture reads 
\begin{eqnarray}
 H_{\text{int}}(t) &=& e^{iH_0t}H_{\text{int}}e^{-iH_0t} \\
 &=& e^{iH_0t}\left[\sigma_z\otimes\sum_{k}\left(g_ka_k^\dagger+g^*_ka_k\right)\right]e^{-iH_0t}, \notag
\end{eqnarray}
where $H_0=H_{\text{S}}+H_{\text{B}}=\frac{\omega_0}{2}\sigma_z + \sum_{k}^{}\omega_ka_k^\dagger a_k$ is the free Hamiltonian of the system and the bath. The corresponding time evolution operator can be computed as
\begin{equation}
U_{\text{SB}}(t) = \mathfrak{T}\exp\left[-i\int_{0}^{t}dt^\prime H_{\text{int}}(t^\prime)\right],\label{Eq:TimeOrder}
\end{equation}
where $\mathfrak{T}$ denotes the time ordering. Formally, \myref{Eq:TimeOrder} can be expanded in Dyson series and usually when the Hamiltonians $H_{\text{int}}(t)$ in different time do not commute, it is difficult to obtain a closed form of the series. In the present scenario it is also true that the Hamiltonians in different time do not commute, but since $[a_k,a_k^\dagger]=1$, it turns out that the commutators of $H_{\text{int}}(t)$ in different time is just a number
\begin{eqnarray}
&&{[H_{\text{int}}(t),H_{\text{int}}(t^{\prime})]} = \sum_{k}2\text{i}|g_k|^2 \sin[\omega_k(t'-t)].
\end{eqnarray} 
This facilitates us to write $U_{\text{SB}}(t)$ as a product of a global time-dependent phase factor and the ordinary (not time-ordered) exponential as follows
\begin{equation}
U_{\text{SB}}(t)=e^{i\chi(t)}\exp\left[-i\int_{0} ^{t} dt^{\prime}H_{\text{int}}(t^{\prime})\right].
\end{equation}
The global phase factor will play no role in the rest of our calculation; therefore, we can avoid it for simplicity and the simple time integration in the argument of the exponential yields
\begin{align}\label{EvOp}
  U_{\text{SB}}(t)&=\exp\left\{\sigma_z\otimes \sum_{k} \left[\xi_{k}(t)a_k^\dagger-\xi_k^*(t)a_k\right]\right\}\nonumber\\
  &= \ket{0}\bra{0}\otimes \mathcal{D}^\dagger  + \ket{1}\bra{1}\otimes \mathcal{D},
\end{align}
with
\begin{eqnarray}\label{xi}
  &&\xi_{k}(t) = \frac{g_k}{\omega_k}\left[1-\exp(\text{i}\omega_kt)\right],\\
  &&\mathcal{D} = \prod_k\exp\left[\xi_{k}(t)a_k^\dagger-\xi_k^*(t)a_k\right].
\end{eqnarray}
Here, $\ket{0}$ and $\ket{1}$ are the ground and excited state of the spin, respectively. Now the system dynamics can be obtained by evolving the total system plus bath and tracing out the bath, which results in the following state of the system at time $t$
\begin{equation}
\rho_{\text{S}}(t)=\begin{pmatrix} 
\rho_{\text{S}}^{00} & e^{-\Gamma(t)}\rho_{\text{S}}^{01} \\
e^{-\Gamma(t)}\rho_{\text{S}}^{10} & \rho_{\text{S}}^{11} 
\end{pmatrix},
\end{equation}
where $e^{-\Gamma(t)} = \text{Tr} (\rho_\text{B} \mathcal{D}^2)$. The  $\Gamma$ is called the decoherence factor which is a function of the interaction strength $g_k$ and frequencies $\omega_k$. Here we can see that the diagonal elements of the density matrix, which represent the populations of the different levels are time-independent and only the off-diagonal elements are decaying over time. This particular form of open quantum system dynamics is called pure dephasing. 

\subsection{Measure of entanglement in spin systems: Concurrence}\label{sec2b}
Entanglement is an important resource for quantum computation and secure communication. It captures the amount of quantum correlations between two quantum systems. For a bipartite quantum system one can define the  entanglement of formation $E_f$ as the minimum amount of entanglement required to construct a correlated state $(\rho)$:
\begin{align}
E_f(\rho) & = \min_{\{\ket{\psi_i},p_i\}}\sum_i p_i E_f(\ket{\psi_i}).
\end{align}
Here, the minimization is carried out over all the decompositions of the density matrix $\rho$
\begin{align}
\rho & = \sum_i p_i \ket{\psi_i}\bra{\psi_i},
\end{align}
such that
\begin{align}
E_f(\ket{\psi_i}) &= S(\rho_A);\quad \rho_A = Tr_B(\ket{\psi_i}\bra{\psi_i}).
\end{align}
Here $S(.)$ is the von-Neumann entropy. For a two-qubit system entanglement of formation is given by
\begin{align}
E_f(\rho) &= h(x)\left(\frac{1+\sqrt{1-\mathcal{C}^2}}{2}\right),
\end{align}
with
\begin{equation}
h(x) = -x\log_2(x) - (1-x)\log_2(1-x).
\end{equation}
The quantity $\mathcal{C}$ is called the concurrence which is calculated as
\begin{align}
\mathcal{C} &= \max(0,\lambda_1-\lambda_2-\lambda_3-\lambda_4),\label{Eq:Con}
\end{align}
where $\lambda_i$'s are the square root of the eigenvalues of the matrix $R$ in descending order, where the matrix $R$ is given by
\begin{align}
R &= \rho (\sigma_y\otimes\sigma_y) \rho^* (\sigma_y\otimes\sigma_y).
\end{align}
Since $E_f$ is a monotonic function of $\mathcal{C}$, it is sufficient to consider $\mathcal{C}$ itself as a measure of entanglement \cite{Wootters}.

{\em Entanglement in a state experiencing local pure dephasing bath.} Now we consider a maximally entangled state $\Ket{\Phi^+} = (\Ket{0}\otimes\Ket{0} + \Ket{1}\otimes\Ket{1})/\sqrt{2}$ of a two-spin system. In this system, one of the spins (say the left one) is under the influence of the pure dephasing dynamics with decoherence parameter $\Gamma(t)$. This will cause the transformation of the state $\Ket{\Phi^+}$ as
\begin{align}
  \Ket{\Phi^+} \to \rho =
  \frac{1}{2}\begin{pmatrix}
    1 & 0 & 0 & \text{e}^{-\Gamma(t)}\\
    0 & 0 & 0 & 0\\
    0 & 0 & 0 & 0\\
    \text{e}^{-\Gamma(t)} & 0 & 0 & 1
  \end{pmatrix}.
\end{align}
The concurrence in the state $\rho$ can be calculated using \myref{Eq:Con}, and we get $\mathcal{C} = e^{-\Gamma(t)}$. Hence, the entanglement in such systems is completely characterized by the decoherence factor $\Gamma$.

\subsection{Non-Hermitian Hamiltonians} \label{sec2c}
The reason that the Hermitian operators are considered observables in quantum mechanics is because they have real spectrum. However, Hermitian operators are only a subset of operators with real eigenvalues. For example, the operators which are symmetric under the combined operation of parity $\mathcal{P}$ and time reversal $\mathcal{T}$, i.~e., which are invariant under the anti-linear operation, $x\rightarrow -x,~p\rightarrow p,~i\rightarrow -i$, admit entirely real spectrum \cite{Bender_Boettcher,PT_Book}. Within the regime of spontaneously unbroken $\mathcal{PT}$-symmetry, one can further introduce a charge conjugation like operator $\mathcal{C}$ to provide a self-consistent description of physical systems in the sense that the time evolution of the $\mathcal{CPT}$ symmetric system becomes unitary \cite{Bender_Brody_Jones}.

One of the simplest ways of constructing the non-Hermitian Hamiltonians $H^{\text{nh}}$ is to add a Hermitian Hamiltonian $H^{\text{h}}$ with a specific form of the non-Hermitian part $H_{\text{n}}(\tau)$ consisting of a free dimensionless parameter, say $\tau$, which characterizes the strength of the non-Hermiticity, i.~e.,
\begin{equation} \label{bath}
H^{\text{nh}}  = H^{\text h} + H_{\text{n}}(\tau).
\end{equation}
The non-Hermitian part $H_{\text{n}}(\tau)$ is chosen in such a way that it becomes $\mathcal{PT}$-symmetric. An alternative way of understanding the non-Hermitian Hamiltonians is to consider the Hamiltonian to be quasi-Hermitian \cite{Scholtz_Geyer_Hahne} or more conveniently pseudo-Hermitian \cite{Mostafazadeh}, where a non-Hermitian Hamiltonian $H^{\text{nh}}$ and a Hermitian Hamiltonian $H^{\text{h}}$ are related by a similarity transformation $H^{\text{h}}=\eta H^{\text{nh}}\eta^{-1}$. As a consequence, the eigenvalues of $H^{\text{nh}}$ belong to the same similarity class as that of $H^{\text{h}}$ with respect to a non-unique, linear, invertible and self-adjoint operator $\eta$ playing the role of the metric. It follows that the non-Hermitian Hamiltonian $H^{\text{nh}}$ becomes Hermitian and the eigenstates of the non-Hermitian Hamiltonian $H^{\text{nh}}$ becomes orthogonal with respect to the so called ``pseudo-inner product", provided that the eigenstates of $H^{\text{nh}}$ and $H^{\text{h}}$ ($|\Phi^{\text{nh}}\rangle$ and $|\Phi^{\text{h}}\rangle$, respectively) are related by the condition $|\Phi^{\text{nh}}\rangle =\eta^{-1}|\Phi^{\text{h}}\rangle$ \cite{Mostafazadeh}. Following this approach, all that requires is to find a suitable metric $\eta$, with respect to which one can describe a non-Hermitian Hamiltonian to be entirely physical.
\section{Non-Hermitian spin-Boson model}\label{sec3}
Our main aim in this article is to introduce a non-Hermitian bath and study its effect on the evolution of a two-level system. We start by replacing the usual harmonic oscillator bath with a non-Hermitian one by following the prescription introduced in Sec.\,\ref{sec2c}. The proposed non-Hermitian bath Hamiltonian in the spin-Boson model reads
\begin{equation} \label{bath2}
H^{\text{nh}}_{\text{B}}  = \sum_k\left(\frac{p_k^{2}}{2m} + \frac{1}{2}k x^{2}  + 2 i \hbar \check{\tau} k p_k x\right).
\end{equation}
with $\omega_k = \sqrt{k/m}$ and $\check{\tau} = \tau/(m \omega_k \hbar)$ having dimension of inverse squared momentum and $\tau$ being dimensionless. The first two terms in \myref{bath2} are the Hamiltonian of the quantum harmonic oscillator in mode $k$ and the last term is the non-Hermitian part which is $\mathcal{PT}$-symmetric. By following the pseudo-Hermiticity approach, we first calculate a metric $\eta=e^{\check{\tau}\sum_k p_k^2}$ that transforms the non-Hermitian bath Hamiltonian \myref{bath2} to a corresponding Hermitian Hamiltonian $H^{\text{h}}_{\text{B}}$ as follows
\begin{eqnarray}\label{bathHer}
H^{\text{h}}_{\text{B}}&=&\eta H^{\text{nh}}_{\text{B}}\eta^{-1}\notag\\ 
&=&\sum_k\left[(1 + 4 \tau^{2})\frac{p_k^{2}}{2 m} + \frac{1}{2}k x^{2} + \hbar \omega_k \tau\right],
\end{eqnarray}
which reduces to a standard harmonic oscillator when we switch off the non-Hermiticity by considering $\tau=0$. Therefore, the parameter $\tau$ carries the signature of the non-Hermiticity and it quantifies the non-Hermiticity in the bath Hamiltonian. Nevertheless, the purpose of all this exercise is to study the effect of non-Hermiticity of the bath Hamiltonian on the dynamical evolution of the spin-system, as well as, to study the role of the non-Hermitian Hamiltonian on the entanglement dynamics of two spin-systems. Considering the bath Hamiltonian $H^{\text{h}}_{\text{B}}$ given in \myref{bathHer}, the new  spin-Boson Hamiltonian for our system reads
\begin{eqnarray}\label{NHSB}
H_{\text{tot}}&&= \frac{\omega_0\sigma_z}{2}+ \sigma_z\otimes\sum_{k}( g_{k}a_k^\dagger+g^{*}_{k}a_k)\notag \\
&&+\sum_{k}\left\{\omega_k\left[a_k^\dagger a_k+\frac{1}{2}+\tau-\tau^2(a_k-a_k^\dagger)^2\right]\right\},
\end{eqnarray}
where the bath Hamiltonian in \myref{NHSB} has been rewritten in terms of the Bosonic ladder operators, with $\hbar=1$. Here we choose to keep the system Hamiltonian $H_{\text{S}}$ same as the usual spin-Boson case. The interaction Hamiltonian $H_{\text{int}}$ also does not change with respect to the usual spin-Boson case, since the potential of the effective Hamiltonian \myref{bathHer} is still in the form of a Harmonic oscillator. 
\subsection{Decoherence due to non-Hermitian bath}
Let us now obtain the dynamics of the spin system in our model by solving it in the interaction picture. Following the steps given in Sec.~\ref{sec2a}, we can directly calculate the $\xi_k(t)$ factor  as
\begin{align}\label{Xi}
 \xi_{k}(t) &=  \frac{8\omega_k\sin^2\frac{\Omega_kt}{2}}{\Omega_{k}^{2}} \left\{\frac{g_k}{4}+\tau^2\text{Re}[g_k]\right\} -\frac{\text{i}g_k\sin(\Omega_k t)}{\Omega_{k}}, 
\end{align}
with $\Omega_k = \omega_k\sqrt{1+4\tau^2}$. Thus, it is straightforward to calculate the operator $\mathcal{D}$ and consequently $\rho_{\text{S}}(t)$, whose elements are given as follows
\begin{equation}
\begin{aligned}
&& \rho_{\text{S}}^{01}(t)=\left[\rho_{\text{S}}^{10}(t)\right]^*=\rho_{\text{S}}^{01}(0)\langle \mathcal{D}^2\rangle, 
\\
&& \rho_{\text{S}}^{00}(t)=\rho_{\text{S}}^{00}(0), \qquad \rho_{\text{S}}^{11}(t)=\rho_{\text{S}}^{11}(0). \label{Matrix2}
\end{aligned}
\end{equation}
The Eq.~\myref{Matrix2} shows that only the off-diagonal elements of the density matrix carries the signature of the decoherence.

The rest of the manuscript deals with the analysis of the decoherence: $e^{-\Gamma(t)}=\langle \mathcal{D}^2\rangle \equiv \text{Tr} \left\{\rho_b \prod_k\exp\left[2\xi_{k}(t)a_k^\dagger-2\xi_k^*(t)a_k\right]\right\}$ which also characterizes the evolution of entanglement for the given system, as indicated in \ref{sec2b}. In what follows, we shall consider a specific case, where we consider the steady state of the bath $\rho_{\text{B}}$ as the thermal state, i.~e.,
\begin{equation}\label{ThEq}
\rho_{\text{B}}(0)=\prod_{k}\rho_{\text{B},k}(T) = \prod_{k}(1-e^{-\beta\omega_{k}})e^{-\beta\omega_{k}a^{\dagger}_{k}a_{k}},
\end{equation}
where $\beta=1/(K_B T)$, with $K_B$ being the Boltzmann constant. The function $\langle \mathcal{D}^2\rangle$ is a well-known entity in quantum optics, which is a product of symmetrically ordered characteristic function for a bunch of harmonic oscillators in thermal equilibrium, and this can be written in a simpler form as
\begin{equation}
\langle \mathcal{D}^2\rangle\equiv e^{-\Gamma(t)}=\prod_k\exp\left[-|\xi_k(t)|^{2}\coth\left({\frac{\omega_{k}}{2 K_B T}}\right)\right].
\end{equation}
This results in
\begin{alignat}{1}\label{Gammat}
\Gamma(t)&=\sum_k\left[\frac{32\tau^2\omega_k\text{Re}[g_k]\text{Im}[g_k]\sin(\Omega_kt)\sin^2(\Omega_kt/2)}{\Omega_k^3}\right. \notag \\
& ~~~\left. +\frac{8\omega^2\sin^4(\Omega_kt/2)\left\{|g_k|^2+8\tau^2\text{Re}[g_k]^2(1+2\tau^2)\right\}}{\Omega_k^4}\right. \notag\\
&~~~\left.+\frac{2|g_k|^2\sin^2(\Omega_kt)}{\Omega_k^2} \right]\coth\left(\frac{\omega_k}{2K_BT}\right),
\end{alignat}
which is the required decoherence factor that will assist our further analysis.
\begin{figure*}
  \subfigure[]{
\includegraphics[width=5cm]{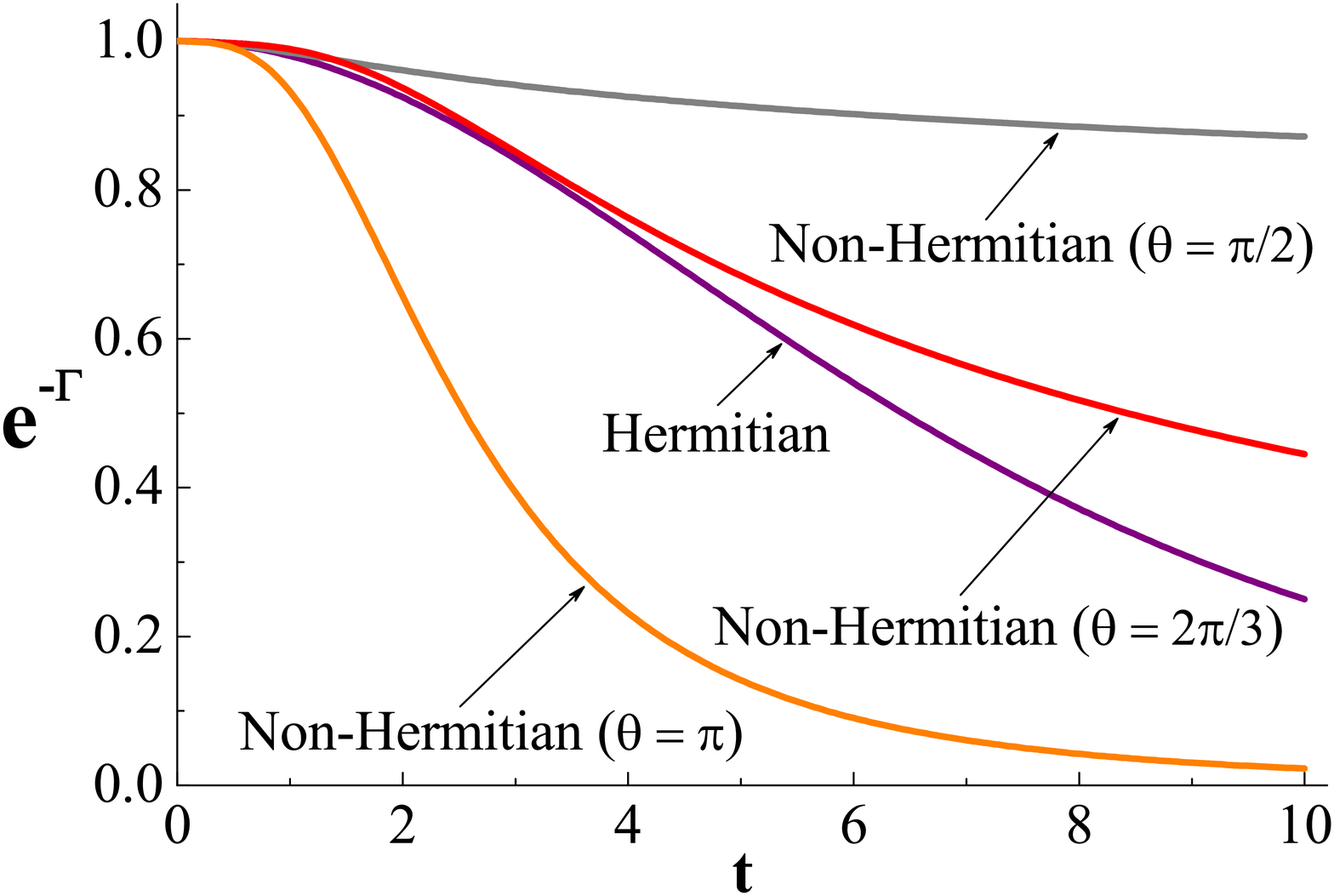}
\label{fig1a}}
  \subfigure[]{
    \includegraphics[width=5cm]{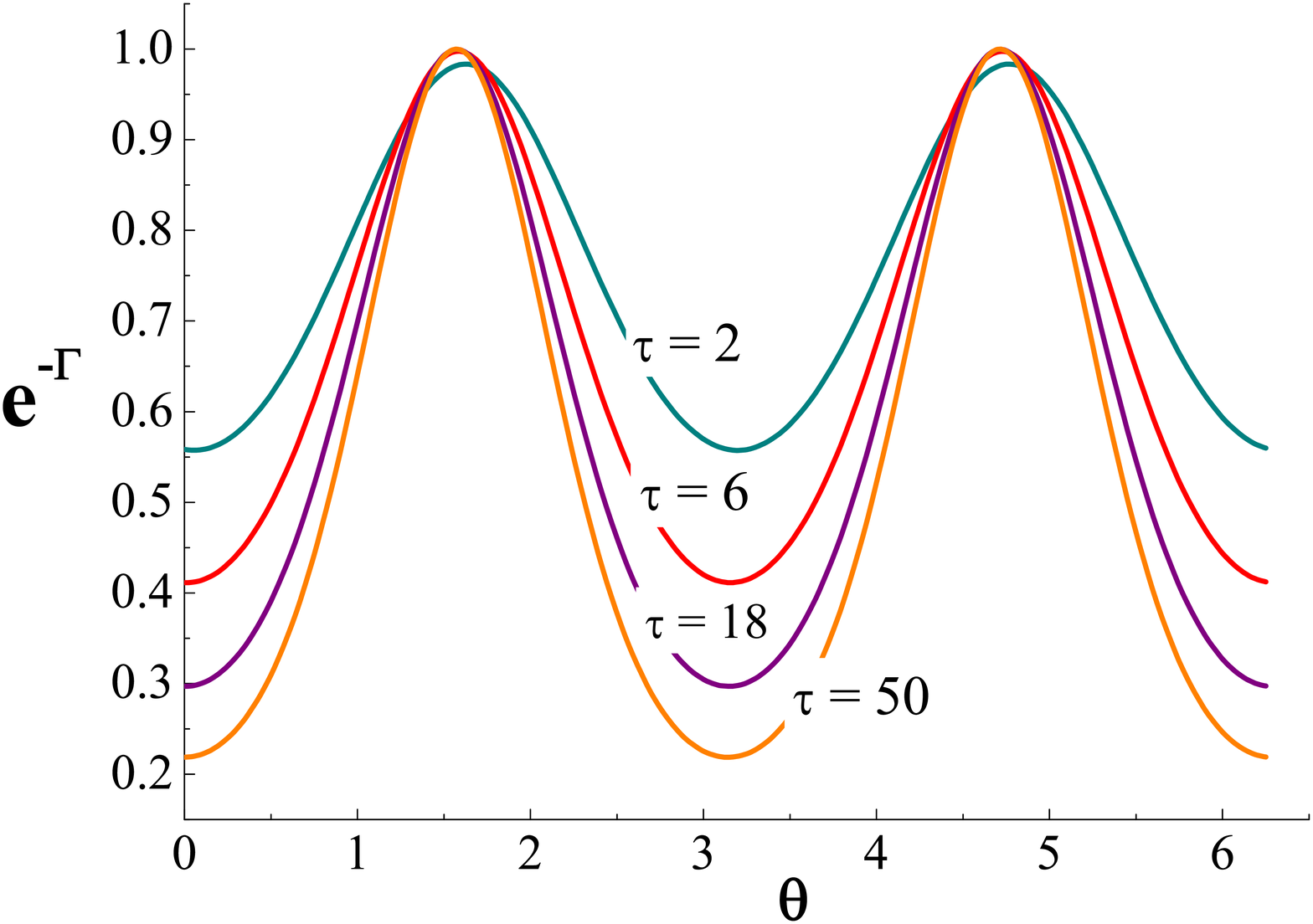}
    \label{fig1b}} 
\subfigure[]{
  \includegraphics[width=5cm]{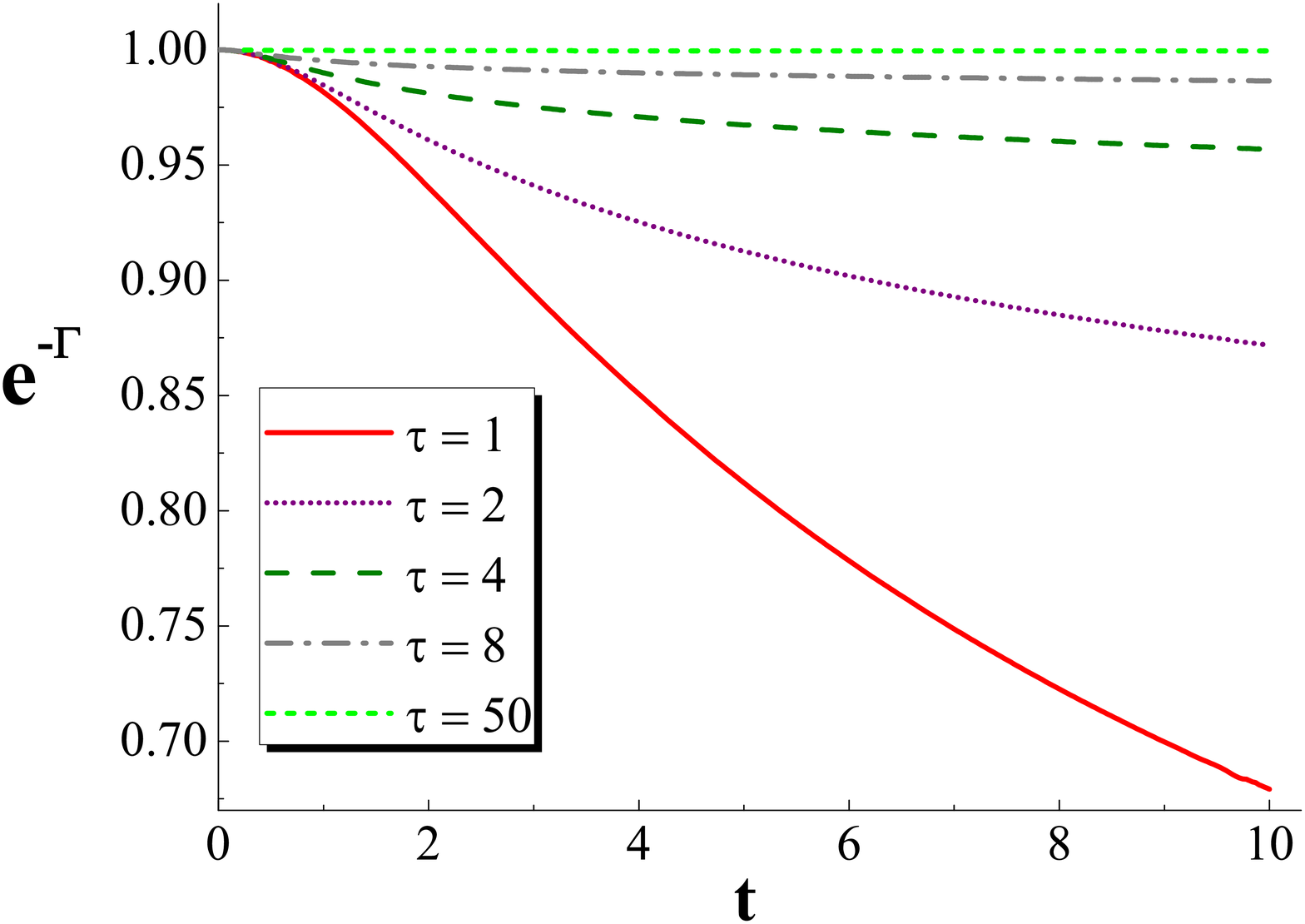}
  \label{fig2}}
\subfigure[]{
  \includegraphics[width=5cm]{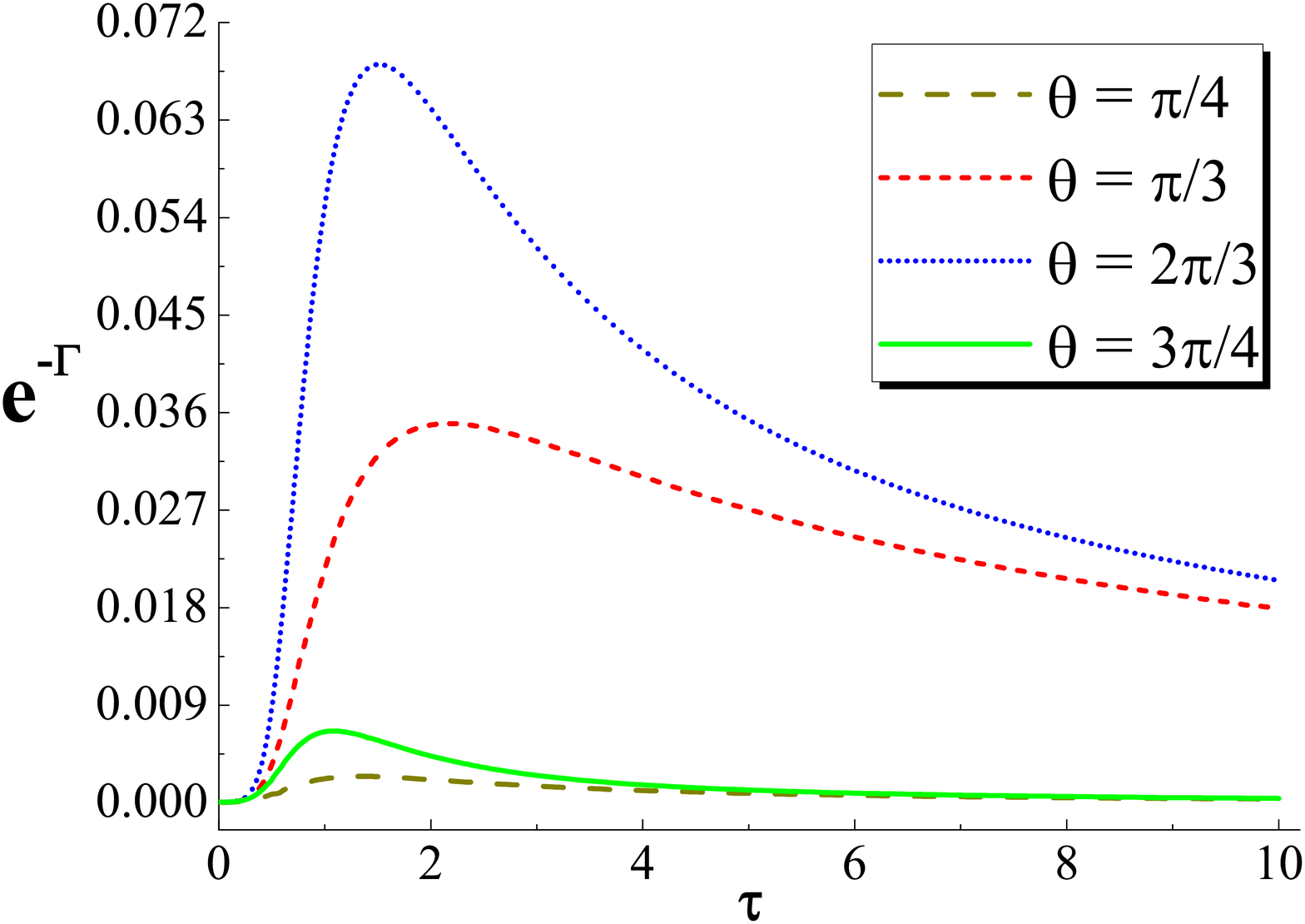}
  \label{fig3a}}
\subfigure[]{
  \includegraphics[width=5cm]{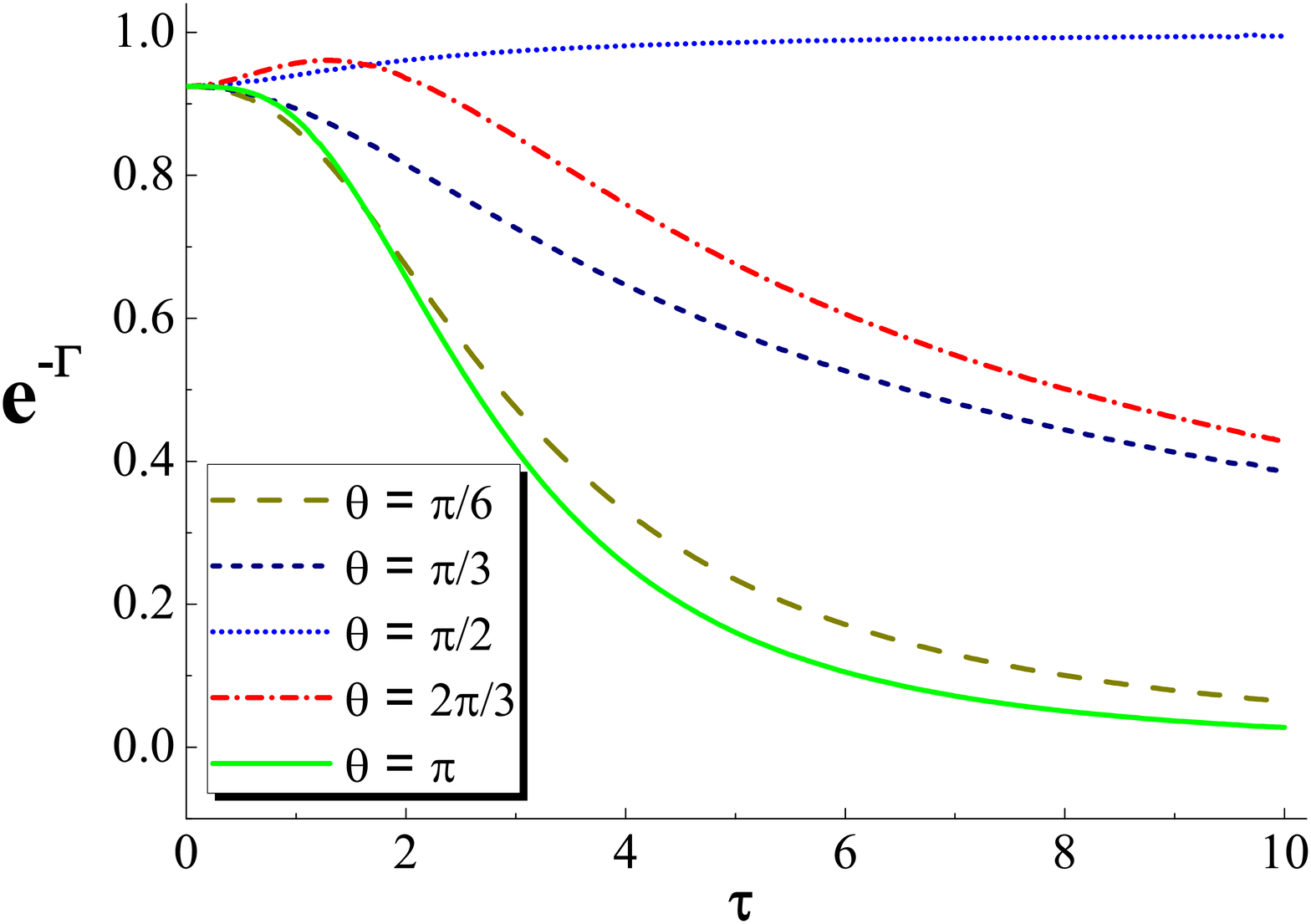}
  \label{fig3b}}
\subfigure[]{
  \includegraphics[width=5cm]{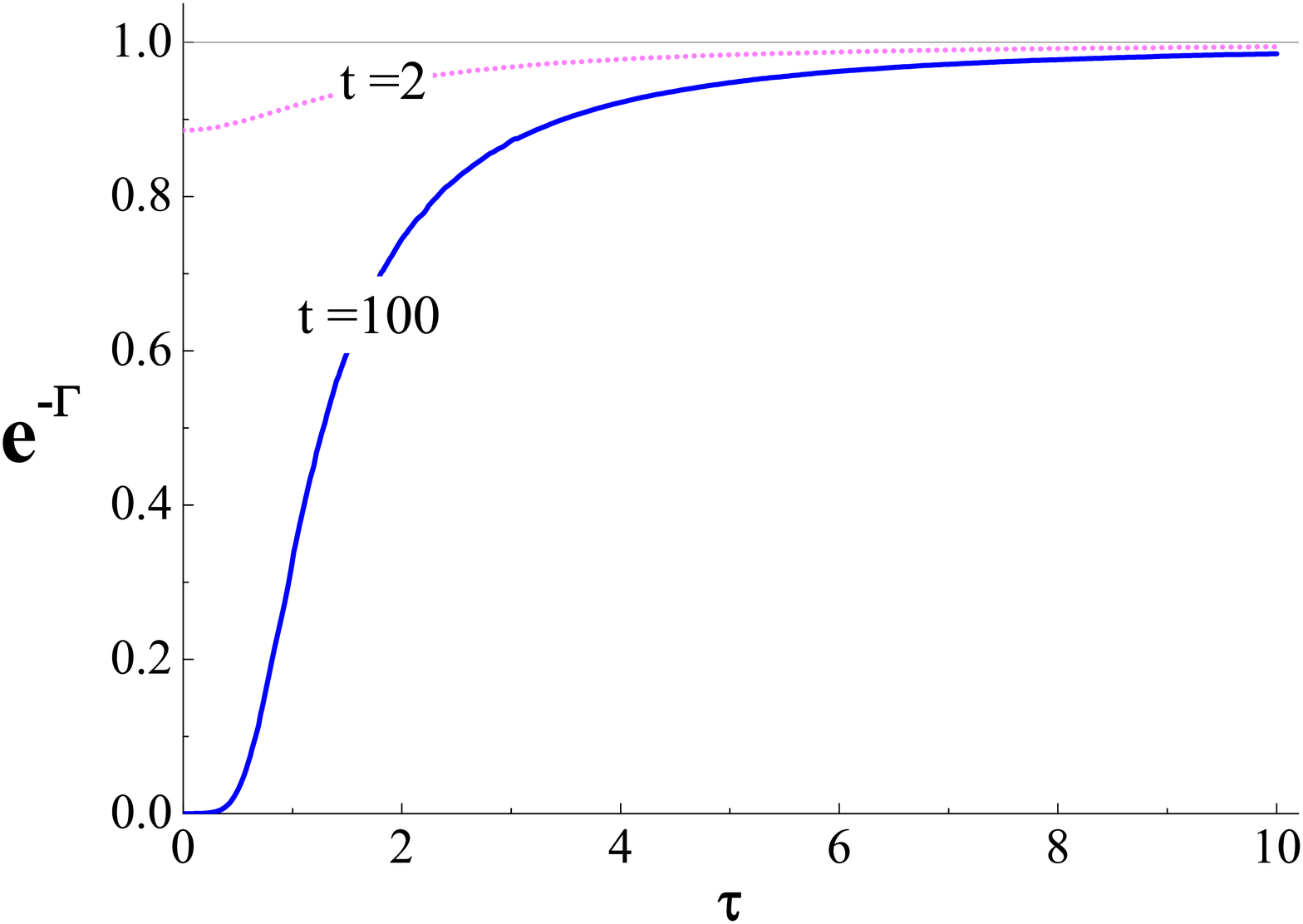}
  \label{fig4}}
 \caption{(Color online) \subref{fig1a} Comparison of the time evolution of decoherence between the standard (Hermitian, $\tau=0$) and non-Hermitian spin-Boson model for different values of $\theta$ with $\tau = 2, A = 1$. \subref{fig1b} Decoherence for non-Hermitian spin-Boson model as the function of $\theta$ for different values of $\tau$, with $A = 0.1, t= 20$. In both panels \subref{fig1a} and \subref{fig1b}, we have used $\omega = 1, T = 300, \Lambda = 0.1$. \subref{fig2} Time evolution of decoherence for different values of $\tau$ for $\theta =\pi/2$ with $\omega=1, T = 300, \Lambda = 0.1, A = 1$. Decoherence for non-Hermitian spin-Boson model as a function of $\tau$ for different values of $\theta$ with $\omega=1, T = 300, \Lambda = 0.1, A = 1$ \subref{fig3a} $t= 120$ \subref{fig3b} $t=2$. \subref{fig4} Decoherence pattern for non-Hermitian model as a function of $\tau$ for small and large time limits for $\theta =\pi/2, \omega = 1, T = 300, \Lambda = 0.1, A = 1$.}
  \label{fig:NH}
\end{figure*}
\subsection{Analysis of the decoherence properties}
In this section, we analyze the decoherence factor that we have obtained in~\myref{Gammat} and study its behavior as a function of the non-Hermiticity parameter $\tau$ and the interaction strength $g_k$. First, we consider that the environment is very large; therefore, we can assume a continuous density of environmental modes such that the sum over the discrete couplings $g_{k}$ can be converted to a continuous description by means of a spectral density $J(\omega)$, i.e.
\begin{equation}
\sum_k|g_k|^2 \rightarrow \int_0^\infty \text{d}\omega J(\omega).
\end{equation}
If we consider the complex coupling constant in the polar form, we can replace $\text{Re}[g_k]^2\rightarrow|g_k|^2\cos^2\theta$, $\text{Im}[g_k]^2\rightarrow|g_k|^2\sin^2\theta$ and $\text{Re}[g_k]\text{Im}[g_k]\rightarrow|g_k|^2\sin\theta\cos\theta$ in \myref{Gammat}. Furthermore, we will consider  spectral density to be linear for sufficiently small frequencies, i.e, $J(\omega)\propto\omega$, and it has a smooth high frequency cutoff quantified by $\Lambda$. Here, let us choose an exponential cutoff of the form $e^{-\omega/\Lambda}$ so that the spectral density takes the form
\begin{equation}
J(\omega)=A\omega e^{-\omega/\Lambda},
\end{equation}
with $A$ being a dimensionless coupling constant. Thus, our assumption implies that the spectral frequency $J(\omega)$ increases approximately linearly for frequencies $\omega<\Lambda$ and decreases for $\omega>\Lambda$, and the decoherence factor $\Gamma(t)$ in \myref{Gammat} becomes
\begin{alignat}{1} \label{GammaNH}
\Gamma(t)&=\int_0^\infty\text{d}\omega \frac{2A\omega e^{-\omega/\Lambda}}{\Omega^4}\left[\Omega^2\sin^2(\Omega t) \right.\\  
&~~~~\left.+16\tau^2\omega\Omega\sin\theta\cos\theta\sin(\Omega t)\sin^2(\Omega t/2)\right. \notag \\
&~~~~\left. +4\omega^2\sin^4(\Omega t/2)\left\{1+8\tau^2(1+2\tau^2)\cos^2\theta\right\}\right] \notag\\
&~~~~\times\coth\left(\frac{\omega}{2K_BT}\right).\notag
\end{alignat}
The Hermitian limit of the decoherence factor can be achieved by setting $\tau = 0$ in the above result \myref{GammaNH}, which results in the well known decoherence factors for the ordinary spin-Boson system as given by~\cite{Schlosshauer_Book}
\begin{equation}\label{GammaH}
\gamma(t)=\int_{0}^{\infty} d\omega  \frac{4 Ae^{-\omega/\Lambda}}{\omega} (1-\cos \omega t)\coth{\left(\frac{\omega}{2 K_{B} T}\right)}.
\end{equation}
An important difference between the Hermitian \myref{GammaH} and non-Hermitian \myref{GammaNH} cases is that in the Hermitian case decoherence factor $\gamma(t)$ does not depend on the phase $\theta$ of the interaction, whereas in the non-Hermitian case, it does. This gives rise to an extra degree of freedom to control the decoherence in our model.

Decoherence in two independent dynamics can be compared by calculating the entanglement in a two-spin system, where the initial state of the system is maximally entangled and only one of the spin is interacting with the bath. In such scenario, the evolution of the entanglement is directly related to the strength of decoherence, as was shown in Sec.~\ref{sec2b}. We shall use the concurrence in such system, i.~e., the factor $\exp[-\Gamma(t)]$ as the measure of the decoherence.

An explicit comparison of the decoherence between the Hermitian and the non-Hermitian cases have been depicted in \ref{fig1a}. It indicates that the decoherence can be slowed down by choosing the parameter $\theta$ appropriately. In our case, $\theta=\pi/2$ gives the best result, whereas $\theta=\pi$ provides the worst result. The $\theta$ dependence of the decoherence is periodic as demonstrated in \ref{fig1b}, which ensures that the study of the decoherence pattern for one cycle in $\theta$ will give us the full information about the pattern. For instance, in \ref{fig1b}, it is obvious that the behavior in between $\theta=0$ and $\theta=\pi$ repeats in the next cycle. It also indicates that when the $\theta$ is in between $0$ and $\pi/2$, the decoherence is slowed down, whereas when the $\theta$ is in between $\pi/2$ and $\pi$ the system decoheres faster.

Apart from the parameter $\theta$, the decoherence pattern depends on two other parameters, namely $\tau$ and the time. Let us first see the dependence on the parameter $\tau$ first. Fig.\,\ref{fig2} shows the time evolution of the decoherence for different values of $\tau$ while keeping the value of $\theta$ constant, i.e. $\theta=\pi/2$. The higher the value of $\tau$ is, the better the decoherence is suppressed, which clearly indicates that by increasing the non-Hermiticity one can slow down the decoherence for a constant value of $\theta$. However, the $\tau$ dependence on the decoherence behavior is not always linear; i.e. the slowing down of decoherence is not proportional to the increase of the parameter $\tau$, which is quite obvious from Fig.\,\ref{fig3a}. As a matter of fact, it is not always true that the decoherence is always slowed down with the increase of the parameter $\tau$, but as it is demonstrated in Fig.\,\ref{fig3a} that there exists a critical value after which the decoherence will be sped up. For instance, consider the value of $\theta=2\pi/3$ in Fig.\,\ref{fig3a}, the decoherence is slowed down when the value of $\tau$ is below $1.3$ (approximately), while it speeds up while $\tau$ is above $1.3$.

However, this analysis is carried out in a relatively large value of time, $t=120$. If we take a small time limit, say $t=2$, the behavior changes completely, which is shown in Fig.\,\ref{fig3b}. Here, we do not see any critical value for which the decoherence will be maximally robust. In short, the decoherence depends on three parameters, $\tau,~ \theta$ and $t$, and one has to choose their values accordingly in order to obtain robust results against the conventional spin-Boson model. However, the outcome of our analysis shows that one can obtain a large set of values for different combinations of $\theta, \tau$ and $t$ for which one can slow down the decoherence. A more interesting behavior follows from Fig.\,\ref{fig4}, where we show the asymptotic behavior of the decoherence for both small and large time as a function of $\tau$ for $\theta=\pi/2$. It indicates that as the non-Hermiticity increases, the decoherence approaches to the maximum robustness asymptotically. For small as well as large time limit, it behaves in an identical way and its asymptotic behavior may be useful to completely avoid decoherence in the system.
\section{Concluding remarks}\label{sec4}
In this article, we have studied a non-Hermitian open quantum system and its decoherence properties by calculating the dynamics of the reduced system in the pure dephasing type. Instead of using a conventional spin-Boson model, which is known to be more vulnerable to decoherence, we explore a non-Hermitian spin-Boson model to control the decoherence effect. We have analyzed the decoherence  by calculating the concurrence in a two-spin maximally entangled system where one of the spin is interacting with the bath. We studied the dependence of decoherence in  three different parameters, namely $\theta,~ \tau$ and $t$, as well as, we have carried out an elaborate discussion on different situations in which the decoherence may be robust against the standard spin-Boson model.

A major difficulty lies in realizing  such a system in a laboratory  where the bath Hamiltonian is non-Hermitian in nature. However, in recent times, the non-Hermitian systems have been realized not only in quantum optics \cite{Guo,Ruter,Xiao}, but also on many other platforms~\cite{Regensburger,Feng,Fleury,Shi,Weimann,Chang,Peng}. Therefore, constructing a non-Hermitian bath with appropriate parameters may be feasible which could provide decoherence free quantum system to be used for quantum information processing.
\section{Acknowledgements} S.\,D.\,acknowledges the support of research grant (DST/INSPIRE/04/2016/001391) from the Department of Science and Technology, Govt.\,of India. A.\,R.\,is supported by a project assistant research fellowship by IISER Mohali. S.\,K.\,G.\,acknowledges the financial support from research grant  ECR/2017/002404 from SERB-DST.


\end{document}